\documentclass[twocolumn,nofootinbib,showpacs,prd]{revtex4-1}%
\usepackage{amsmath}
\usepackage{amsfonts}
\usepackage{amssymb}
\usepackage{graphicx}

\begin{document}

\title{Kinematic and statistical inconsistencies of Ho\v rava-Lifshitz cosmology}

\author{Orlando Luongo}	
\email{orlando.luongo@lnf.infn.it}
\affiliation{Istituto Nazionale di Fisica Nucleare, Laboratori Nazionali di Frascati, 00044 Frascati, Italy}
\affiliation{School of Science and Technology, University of Camerino, I-62032, Camerino, Italy}
\affiliation{Instituto de Ciencias Nucleares, Universidad Nacional Aut\'onoma de M\'exico, AP 70543, M\'exico DF 04510, Mexico}

\author{Marco Muccino}
\email{marco.muccino@lnf.infn.it}
\affiliation{Istituto Nazionale di Fisica Nucleare, Laboratori Nazionali di Frascati, 00044 Frascati, Italy}

\author{Hernando Quevedo}
\email{quevedo@nucleares.unam.mx}
\affiliation{Instituto de Ciencias Nucleares, Universidad Nacional Aut\'onoma de M\'exico, AP 70543, Ciudad de M\'exico 04510, Mexico}
\affiliation{Dipartimento di Fisica and ICRA, Universit\`a di Roma "Sapienza", P.le Aldo Moro 5, I-00185, Roma, Italy}
\affiliation{Institute of Experimental and Theoretical Physics,
	Al-Farabi Kazakh National University, Almaty 050040, Kazakhstan}

\date{\today}

\begin{abstract}
We investigate the validity of a \emph{minimal} cosmological model derived from the renormalizable Ho$\check{\textrm{r}}$ava action at low redshift scales by using different cosmological and statistical tests. Assuming pure attractive gravity, i.e., $\lambda>1/3$ in the Ho\v rava action, we compare the Union 2.1 supernova type Ia data with the kinematics following from a model-independent approach.
The two approaches, although compatible, lead to explicit cosmographic constraints on the free parameters of the Ho\v rava action, which turn out to be in strong disagreement with the $\Lambda$CDM, $w$CDM and Chevallier-Polarski-Linder scenarios. To show this, we use standard diagnostic tools of regression models, namely the Akaike  and the Bayesian Information Criteria. Using such model-independent statistical methods, we show that  Ho$\check{\textrm{r}}$ava-Lifshitz cosmology differs from the standard dark energy scenarios, \emph{independently} of the number of free parameters involved in the model.  Since this result is valid at small redshift domains, it indicates the presence of inconsistencies in the minimal version of Ho$\check{\textrm{r}}$ava-Lifshitz cosmology even at the level of background cosmology.
\end{abstract}


\maketitle


\section{Introduction}

Even though general relativity (GR) and quantum mechanics are considered  cornerstones of modern physics, all the attempts to formulate a \emph{unified theory} of quantum gravity have been so far unsuccessful \cite{citazione1}. Interesting technical results have been obtained in different approaches \cite{citazione2}, albeit the physical problem remains still open due, in part, to the fact that GR is non-renormalizable in the ultraviolet (UV) regime \cite{nonren}. Recently, Ho\v rava \cite{Horawa} proposed a model that partly solves the problem at the UV limit.

\noindent The model is renormalizable \cite{Orla} and non-relativistic in the UV regime. Further, it reduces to Einstein's GR with a non-vanishing cosmological constant at the infrared (IR) limit. In the Ho$\check{\textrm{r}}$ava picture, space and time show up with different scalings at the UV fixed point, i.e., $x^i \to  l x^i,\ t \to l^z t$, with $z$ representing the scaling exponent. This model is now commonly known as \emph{Ho$\check{\textrm{r}}$ava-Lifshitz} (HL) theory \cite{edue} and turns out to be renormalizable for $z = 3$. The original model
had the complication that the Schwarzschild-AdS black hole solution was not recovered at the IR limit \cite{BlackHoles}. This caveat was healed by introducing an additional parameter, which modifies the IR behavior \cite{GeneralizzazioneHorawa,adddreee} and  leads  to the generalized Ho\v rava action:
\begin{align}
\nonumber
S_g = \int &\left[\frac{2}{\kappa^2}\left(K_{ij}K^{ij}-\lambda K^2\right)-\frac{\kappa^2}{2\nu^4}C_{ij}C^{ij}+\right.\\
\nonumber
&\ \left.\frac{\kappa^2\mu}{2\nu^2}\epsilon^{ijk} R^{ }_{i\ell} \nabla_{j}R^{ \ell}{}_k -\frac{\kappa^2\mu^2}{8} R^{ }_{ij} R^{ ij} +\right.\\
\nonumber
&\ \left. \frac{\kappa^2 \mu^2}{8(3\lambda-1)}
\left(\frac{4\lambda-1}{4}R^2-\Lambda_W R+3\Lambda_W^2\right)+\right.\\
\label{azione}
&\ \left. \frac{\kappa^2 \mu^2\omega}{8(3\lambda-1)} R\right]N \sqrt{g} d^4 x\,,
\end{align}
where
\begin{align}
K_{ij}&=\frac{1}{2N} (\dot g_{ij}-\nabla_i N_j -\nabla_j N_i)\,,\\
C^{ij}&= \epsilon^{ikl} \nabla_k \left(R^j_l -\frac{1}{4} R \delta^j_l\right)\,,
\end{align}
are the extrinsic curvature and the Cotton tensor respectively, The dot ``$\,\dot{}\,$'' represents the derivative with respect to the time coordinate, and $R$ is the scalar curvature. Further, $N$ and $N_i$ are the lapse and shift in the 3+1 decomposition, i.e.
$ds^2 = -N^2 c^2 dt^2 + g_{ij}(dx^i + N^idt)(dx^j+N^jdt)$.

\noindent Although appealing, the HL model presents a statistical drawback, i.e., it depends upon six parameters:
$\kappa$, $\lambda$, $\mu$, $\nu$, $\Lambda_W$ and $\omega$, which are not completely free. In fact, they determine the speed of light $c$, the gravitational constant $G$ and the cosmological constant $\Lambda$ by means of the relations \cite{Muinpark}:
\begin{equation}
\label{uno}
c^2 = \frac{\kappa^4\mu^2|\Lambda_W|}{8(3\lambda-1)^2}\ ,\
G=\frac{\kappa^2c^2}{16\pi(3\lambda-1)}\ ,\
\Lambda_W=\frac{2}{3}\Lambda\,.
\end{equation}
{Notice that our notation is different from the original one used by Ho$\check{\textrm{r}}$ava \cite{Horawa}}
\begin{itemize}
  \item $G_{here}=G_{Horava}/c^3$,
  \item $\Lambda_{here} = \Lambda_{Horava}$,
\end{itemize}
so that dimensionally  $[G] = L^3/MT^2$, $[c]=L/T$, and $[\Lambda] = [\Lambda_W]=1/L^2$. In addition,
\begin{itemize}
  \item $\lambda<1/3$ implies the presence of repulsive gravitational effects, and
  \item $\lambda>1/3$ leads to attractive gravity.
\end{itemize}
To be in agreement with observations, we limit ourselves to the second case. Consequently, the HL model reduces its complexity as it  possesses only three free parameters, which must be chosen in accordance with observations. In the case of cosmology, for  instance, the three free parameters make the HL model comparable with the Chevallier-Polarski-Linder (CPL) model \cite{cipielle}.

The main purpose of this paper is to revise the HL cosmology by using cosmic kinematics and statistical tests, i.e., selection criteria for  the free constants entering the HL model in a homogeneous and isotropic background. We test the corresponding modified Friedmann equations by using cosmography, which consists in a model-independent kinematical procedure acting on the Taylor series of the luminosity distance
at $z\ll1$. We compare these results with cosmological limits on the free parameters of HL cosmology, considering the conditions valid at
the IR limit. We thus explore the behavior of the HL model at low redshift scales by  comparing it with a few cosmological models, namely, the standard $\Lambda$CDM paradigm, and the $\omega$CDM and CPL scenarios. By using cosmography and statistics, we will find inconsistencies of the HL model with respect to the standard cosmological model.

The article is divided as follows. In Sec. II, we provide a short overview of the main features of HL cosmology. We also present the modified Friedmann equations and the evolution equation for the Hubble rate. In Sec. III, we explore the simple case of a dust-dominated universe and present explicit solutions for the scale factor and the Hubble parameter. These solutions are used to compare  the HL model with cosmic data, in particular, with the Union 2.1 data survey of type Ia supernovae.
In Sec. IV, we consider the cosmographic representation of HL cosmology and find the corresponding bounds determined by kinematics. We us the value of the cosmographic parameters to find bounds on the free parameters of the HL model.
In Sec. V, we employ both cosmological and kinematic tests determined in the previous sections, to establish two statistic information criteria. Finally, in Sec. VI, we point out the inconsistencies of the HL model at the level of background cosmology.


\section{Brief overview of Ho\v rava-Lifshitz cosmology}

Several possibilities are available to describe the homogeneous and isotropic universe in the framework of HL gravity. Here, we start from the Friedmann-Lema\^itre-Robertson-Walker (FLRW) metric and replace it in Eq.($\ref{azione}$) to obtain a reduced action from which the corresponding fiel equations can be obtained. First, we notice that the free parameters entering the HL model should respect all the current bounds at local and cosmic scales.
In particular, for the set of parameters $\kappa$, $\lambda$, $\mu$, and $|\Lambda_W|$, we get from Eq.(\ref{uno}) that
$\kappa^2=16\pi(3\lambda-1)G/c^2$ and $|\Lambda_W|\mu^2=c^6/(32\pi^2 G^2)$. Then,
by using the most recent values $G=6.67\times 10^{-11}$~N\,m$^2$kg$^{-2}$ and $c=299792458$~m/s from \cite{data}, we infer that
\begin{align}
\label{nulla2}
\kappa &= 1.93\sqrt{3\lambda-1}\times  10^{-13} \left({\rm m\,kg}^{-1}\right)^{\frac{1}{2}}\,,\\
\label{nulla22}
|\Lambda_W|\mu^2 &= 5.16\times \,10^{68}\left({\rm kg\,s}^{-1}\right)^2\,.
\end{align}
Eqs.~\eqref{nulla2} and \eqref{nulla22} represent the numerical bounds for the HL model to be compatible with Einstein's gravity in the corresponding limit \cite{bronnikov1988,Melnikov2009,Ivashchuk2014,2013SoSyR..47..386P}

\noindent To compute the modified Friedmann equations in the HL framework,  we are forced to perform the analytic continuation
$\mu^2 \to -\mu^2$ in
the original action, so that the upper (lower) sign corresponds to
the AdS, $\Lambda_W < 0$, (dS, $\Lambda_W > 0$) case. We find that the case $\Lambda_W > 0$ is the only one to be favored by local scale limits, exhibiting an extra fine-tuning in the HL scenario.

\noindent Under the hypothesis of a perfect-fluid source with energy density and pressure $\rho$ and $p=w\rho$, respectively, we get the modified Friedmann equations \cite{Muinpark}:
\begin{align}\label{hu}
\frac{\dot{a}^2}{a^2}&=b_1\Bigg\{\rho \pm b_2 \left[-\Lambda^2_W + \frac{2k(\Lambda_W-\omega)}{r^2_0a^2}-\frac{k^2}{r^4_0a^4}\right]\Bigg\}\,,\\
\frac{\ddot a}{a}&=b_1\left[-\frac{1}{2}(\rho+3p) \pm b_2 \left(-\Lambda^2_W+\frac{k^2}{r^4_0a^4}\right)\right]\,,\nonumber
\end{align}
with $b_1=\kappa^2/[6(3\lambda-1)]$ and $b_2=3\kappa^2\mu^2/[8(3\lambda-1)]$. We see that the free parameter $\nu$ does not appear in the field equations of the FLRW model. Thus, any cosmological test by itself is not enough to completely fix the freedom of the parameters,
entering the generalized Ho$\check{\textrm{r}}$ava action (\ref{azione}).

\noindent The corresponding differential equation, obtained by combining the above two expressions, can be recast as
\begin{widetext}
\begin{equation}
\label{Friedman}
\dot H +\frac{3}{2}(1+w)H^2 -\frac{\Lambda}{2}(1+w)c^2 + \frac{\Lambda k c^2}{3\Lambda_W^2r_0^2a^2} \left[(1+3w)(\Lambda_W-\omega)+\frac{k}{2r_0^2a^2}(1-3w)\right]
=0\,,
\end{equation}
\end{widetext}
which, in turn,  can be rewritten in terms of the more convenient redshift variable $z$ as:
\begin{widetext}
\begin{equation}
\label{allilbrodo}
(1+z)\frac{dH^2}{dz}-3(1+w)H^2+\Lambda(1+w)c^2-\frac{\Lambda kc^2}{3\Lambda_{W}^{2}r_{0}^{2}}\Big[(1+3w)(\Lambda_W-\omega)(1+z)^2+\frac{k}{2r_{0}^{2}}(1-3w)(1+z)^4\Big]=0\,.
\end{equation}
\end{widetext}
Here we defined the Hubble rate $H\equiv \dot{a}/a$, the redshift  $a=(1+z)^{-1}$, and $|\Lambda| = 3\mu^2\kappa^4\Lambda^2_W /16c^2(3\lambda-1)^2$.

\noindent A direct comparison with Einstein's theory shows that the term proportional to $a^{-4}$ represents the contribution from higher-derivative terms present in the generalized HL model, giving a correction to the pure radiation case.
This term vanishes for $k=0$ or $w=1/3$. Consequently, in the case of a flat universe (with arbitrary equation of state) and a
radiation dominated universe only (with arbitrary spatial curvature), it is not possible to differentiate the HL model from Einstein's
gravity. For our purpose, we limit our attention to the lowest redshift regime, corresponding to our observable universe,  neglect
the contribution due to radiation,  and assume a non-flat universe.
Therefore, to complete this scheme, it is necessary to characterize the typology of fluid to be accounted in the HL cosmology.
In what follows, we will consider the case  of dust, which is the simplest case of a matter term with an equation of
state (EoS) of the form $p=0 \ (w=0)$.


\section{Numerical bounds for the dust-like case}

In the framework of HL gravity, the fluid driving the observed speed up is contained \emph{inside} the model.
This means that the simplest scenario to be considered in connection with
Eq.~\eqref{allilbrodo} consists of pressureles matter\footnote{For a different perspective on matter with pressure, see \cite{luongomuccino}.}. To this end, one makes use of cold dark matter and baryons, which do not contribute to accelerate the universe today,
leaving $\Lambda_W$ alone to be responsible for the observed dynamics. Moreover, we found previously the Ho$\check{\textrm{r}}$ava corrections to the Friedmann equations, showing that they become relevant for different regimes. For example, at the low redshift regime, the term $\omega$ does not contribute so that it is difficult to bound it in a  FLRW universe. In order to fix the constraints by using cosmography, one of the first steps consists in finding  the function $H(z)$ by solving the differential equation (\ref{allilbrodo}). Thence, by assuming a constant barotropic factor $w$ for a given cosmological fluid, $H(z)$ can be written as
\begin{align}
\nonumber
E^2(z)=&\sum_i\Omega_i(1+z)^{3(1+w_i)}+\Omega_k\left(1-\frac{\Omega_\omega}{\Omega_\Lambda}\right)(1+z)^2+\\
\label{hdizeta}
&\Omega_{r}^{*}(1+z)^4+\Omega_\Lambda\,,
\end{align}
where $E^2(z)=H^2/H_0^2$ and $|\Omega_\omega|\equiv\omega c^2/(2H_0^2)$. Note that the main difference with respect to
 $\Lambda$CDM is the correction $(1-\Omega_\omega/\Omega_\Lambda)$.
In addition, the sum over all the possible contributions shows that, when $w_i$ is due to radiation, we can replace the radiation counterpart by
$\Omega_{r}^{*}\equiv\Omega_k^2/(4\Omega_\Lambda)+\Omega_{r}$.

\noindent As a particular case, the Friedmann equations in Einstein's theory predict $\dot H+3/2(1+w)H^2-\Lambda/2(1+w)c^2$=0 for a flat universe.
The consequences of the special cases $k=0$ or $w=1/3$ are analyzed by using the well--known standard solutions of
$\Lambda$CDM
\begin{equation}
\label{at}
a(t)=a_0
\left[F\left(\frac{c\sqrt{3\Lambda}}{2}(1+w)t+t_i\right)\right]^{\frac{2}{3(1+w)}}\,,
\end{equation}
where $F(x)=\sinh(x)$ if $\Lambda>0$ and $F(x)=\sin(x)$ if $\Lambda<0$. In our case, $\Lambda_W>0$ implying that $\Lambda>0$, so that:
\begin{equation}\label{h}
H(t)=c\sqrt{\frac{\Lambda}{3}}\coth\Big[\frac{c\sqrt{3\Lambda}}{2}(1+w)t+t_i\Big]\,.
\end{equation}
For $w=0$, we can rewrite Eq. ($\ref{hdizeta}$) as
\begin{align}
\nonumber
E^2(z)=&\Omega_m(1+z)^{3}+\Omega_k\left(1-\frac{\Omega_\omega}{\Omega_\Lambda}\right)(1+z)^2+\\
\label{hdizeta2}
&\Omega_{r}^{*}(1+z)^4+\Omega_\Lambda\,,
\end{align}
which under the assumption that $E(0)=1$ implies the constraint
\begin{equation}
\label{jew}
\Omega_{m}+\Omega_{\Lambda}+\Omega_{k}\left(1-\frac{\Omega_\omega}{\Omega_\Lambda}\right)+\frac{\Omega_{k}^{2}}{4\Omega_\Lambda}+\Omega_r=1\,.
\end{equation}

\noindent We can now perform the cosmological test, taking into account the above results.
Numerical priors of compatibility for $\Omega_\Lambda$, $\Omega_{\Lambda_W}$, and $\mu$ can be obtained by comparing the results of Ref.~\cite{Planck2018} and Ref.~\cite{Riess2016}.\footnote{Ref.~\cite{Riess2016} gets only  $H_0$. For the other cosmological parameters, we considered the values obtained from the analysis of Type Ia supernovae, i.e., $\Omega_m=0.271\pm0.012$ and $\Omega_\Lambda=0.729\pm0.012$ \cite{Suzuki2012}.}
This comparison is necessary in view of the recent $3.4$--$\sigma$ tension arisen between the value of the Hubble constant $H_0=(67.66\pm0.42)$~km\,s$^{-1}$Mpc$^{-1}$, inferred from the Cosmic Microwave Background (CMB) power spectra \cite{Planck2018}, and the value of $H_0=(73.24\pm1.74)$~km\,s$^{-1}$Mpc$^{-1}$ inferred from new, near-infrared observations of Cepheid variables in $11$ host galaxies of recent type Ia supernovae (SNe Ia) \cite{Riess2016}. The priors are summarized in Tab.~\ref{table1}.
\begin{table}[t]
\label{table1}
\begin{center}
\begin{tabular}{l|c|c}
\hline\hline
Parameter 			 			& CMB 								& SNIa \\
\hline
$\Omega_\Lambda$			& $0.6889\pm0.0056$		&	$0.729\pm0.012$ \\
$\Omega_{\Lambda_W}$	&	$0.4593\pm0.0037$		& $0.486\pm0.008$ \\
$\mu/(10^{60}$~kg\,m\,s$^{-1}$) 				& $2.646\pm0.027$			&	$2.376\pm0.069$ \\
\hline
\end{tabular}
\caption{Numerical priors obtained from $\Omega_{\Lambda_W}=2\Omega_\Lambda/3$ with the values of $H_0$ from CMB \cite{Planck2018} and SNe Ia data \cite{Riess2016}.}
\end{center}
\end{table}

\begin{figure}
\centering
\includegraphics[width=0.9\hsize,clip]{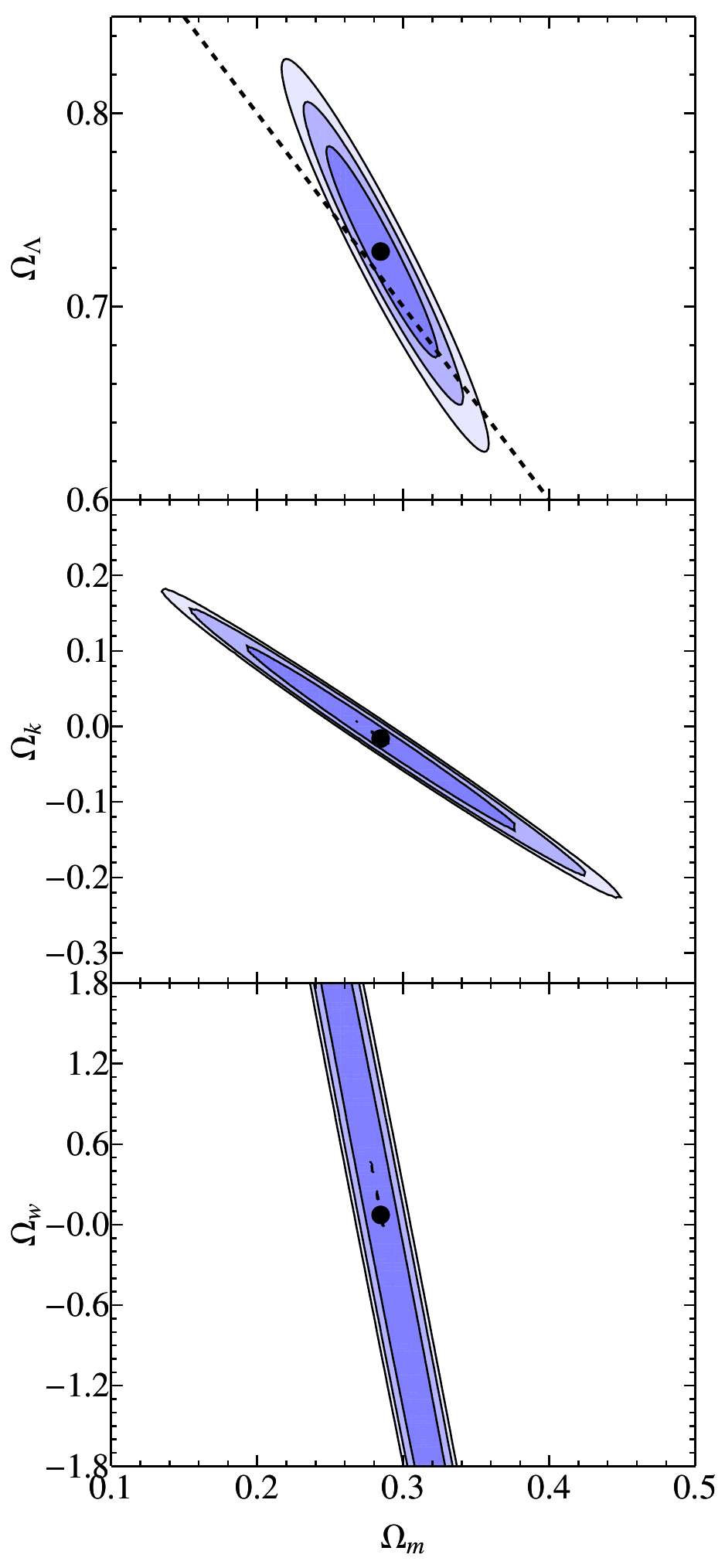}
\caption{$\Omega_\Lambda$--$\Omega_m$ (top panel), $\Omega_k$--$\Omega_m$ (middle panel), and $\Omega_\omega$--$\Omega_m$ (bottom panel) contours of Ho$\check{\textrm{r}}$ava model. The black dots mark the best-fit values and the blue shaded areas show the $1$, $2$, and $3$--$\sigma$ confidence regions (from the darker to the lighter). The dashed line in the top panel marks the allowed values of $\Omega_\Lambda$ and $\Omega_m$ for a flat universe.}
\label{figura2}
\end{figure}

\noindent We now constrain the model parameters by fitting the observational SNe Ia data from the most recent Union $2.1$ compilation
\cite{Suzuki2012}. Since SNe Ia can be considered as standard candles,
they play the role of distance indicators.
The corresponding luminosity distance
\begin{equation}
d_L(z) =  \frac{c}{H_0} (1+z) \int_0^z \frac{dz^\prime}{E(z^\prime)}
\end{equation}
contains information on the cosmological parameters and can be expressed through the distance modulus $\mu_{\rm SN}$
\begin{equation}
\mu_{\rm SN} = 25 + 5 \log_{10} \left(\frac{d_L}{\rm Mpc}\right)\,,
\end{equation}
with the error $\sigma\mu_{\rm SN}$ including the SNe Ia systematics.
\begin{figure}
\centering
\includegraphics[width=\hsize,clip]{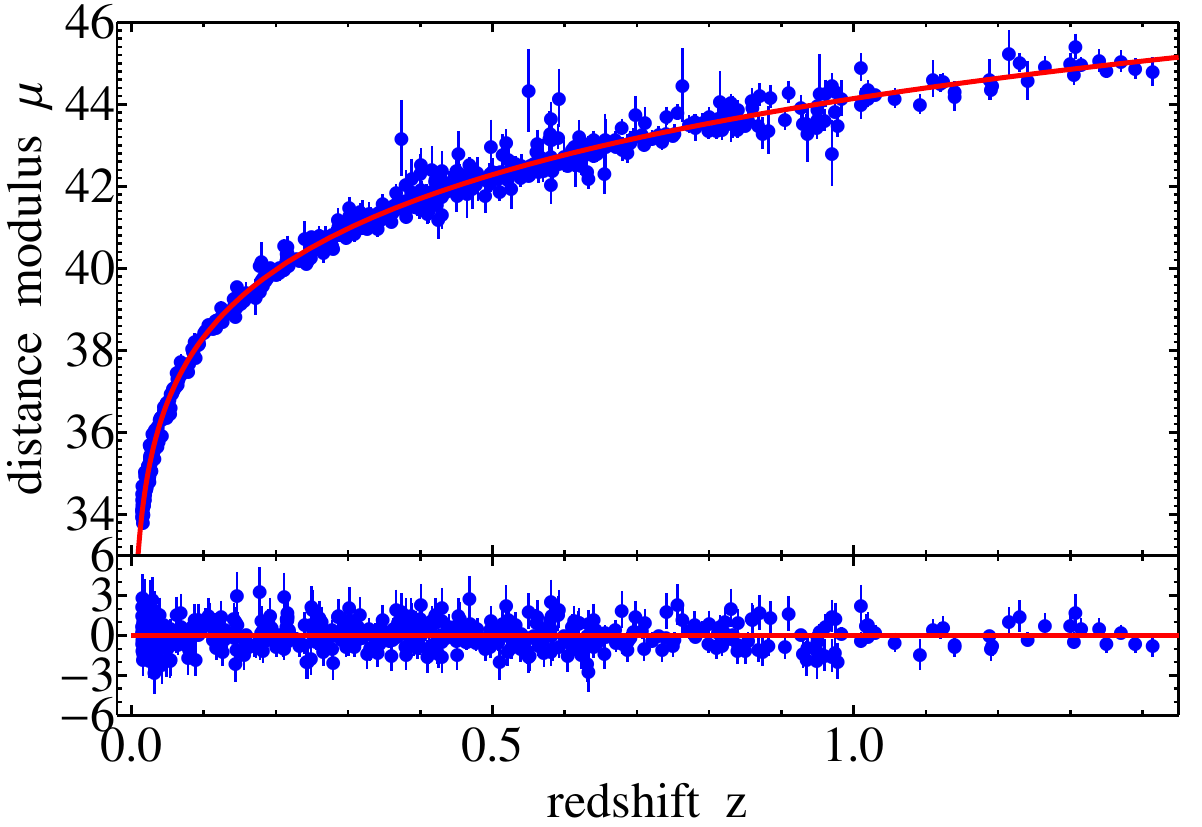}
\caption{The SNIa distance moduli $\mu_{\rm SN}$ distribution with the redshift $z$ and their corresponding errors $\sigma\mu_{\rm SN}$. The superposed red line is obtained by using the fiducial values $H_0=70$~km\,s$^{-1}$Mpc$^{-1}$, $\Omega_m=0.27$ and $\Omega_\Lambda=0.73$.}
\label{figura1}
\end{figure}
To constrain the free parameters of the model, one has to compare the observed $\mu_{\rm SN}$ from SNe Ia (see Fig.~\ref{figura1}) with the ones within the proposed model.
This can be done through the $\chi$-square statistic
\begin{equation}
\chi^{2} =
\sum_{i=1}^{N}\frac{(\mu_{\rm SN,i}^{\mathrm{theo}}-\mu_{\rm SN,i}^{\mathrm{obs}})^{2}}{\sigma\mu_{\rm SN,i}^{2}}\,.
\end{equation}
The best-fit parameters for the Ho$\check{\textrm{r}}$ava model, which minimize the quantity $\chi^{2}$, are summarized in Tab.~\ref{table200} and the corresponding contour plots are shown in Fig.~\ref{figura2}.\footnote{In all fits, the Hubble constant has been initially set free. However, for all the models explored in this work, it has always been assumed that  $H_0=70$~km\,s$^{-1}$Mpc$^{-1}$.}
\begin{table}[t]
\label{table200}
\centering
\begin{tabular}{c|c|c|c}
\hline\hline
\multicolumn{4}{c}{Best-fit parameters ($\chi^2$/DOF$=562.23/576$)}\\
\hline
$\Omega_m$						&	$\Omega_{\Lambda}$		& $\Omega_k$    										& $\Omega_\omega$	\\
\hline
$0.285\pm0.094$				&	$0.729\pm0.042$  			& $-0.02\pm0.12$										& $0.08\pm3.68$		\\
\hline\hline
\multicolumn{4}{c}{Derived quantities}\\
\hline
$\Omega_0$						&	$\Omega_{\Lambda_w}$	& $\mu/(10^{60}$kg\,m\,s$^{-1}$)   & $\omega/(10^{-54}$m$^{-2}$) \\
\hline
$1.01\pm0.10$				  & $0.486\pm0.028$				& $2.488\pm0.072$   								& $9.0\pm422.0$   						 \\
\hline
\end{tabular}
\caption{Top part: Best-fit values of $\Omega_m$, $\Omega_{\Lambda}$, $\Omega_k$, and $\Omega_\omega$ for Ho$\check{\textrm{r}}$ava model and the value of the $\chi^2$ statistic test over the number of degrees of freedom (DOF). Bottom part: Summary of the quantities derived from the above best-fit parameters, i.e., the total density $\Omega_0$, $\Omega_{\Lambda_W}$, $\mu$, and $\omega$.}
\end{table}

\noindent In the next section, we discuss the role of cosmography as a tool to determine if the results of Tab.~\ref{table200} are in agreement with the kinematics predicted by the HL model.


\section{Kinematics of Ho\v rava-Lifshitz cosmology}

All cosmic tests are commonly based on assuming that the underlying model represents the best suit to fit cosmic data surveys. This caveat introduces some sort of  model-dependence in the fitting procedure \cite{orlando1,orlando2,orlando3}. It is therefore relevant to use \emph{model-independent} treatments to establish the paradigms, which better work among the wide number of possibilities \cite{peter}.

\begin{figure*}[t]
\centering
\includegraphics[width=0.75\hsize,clip]{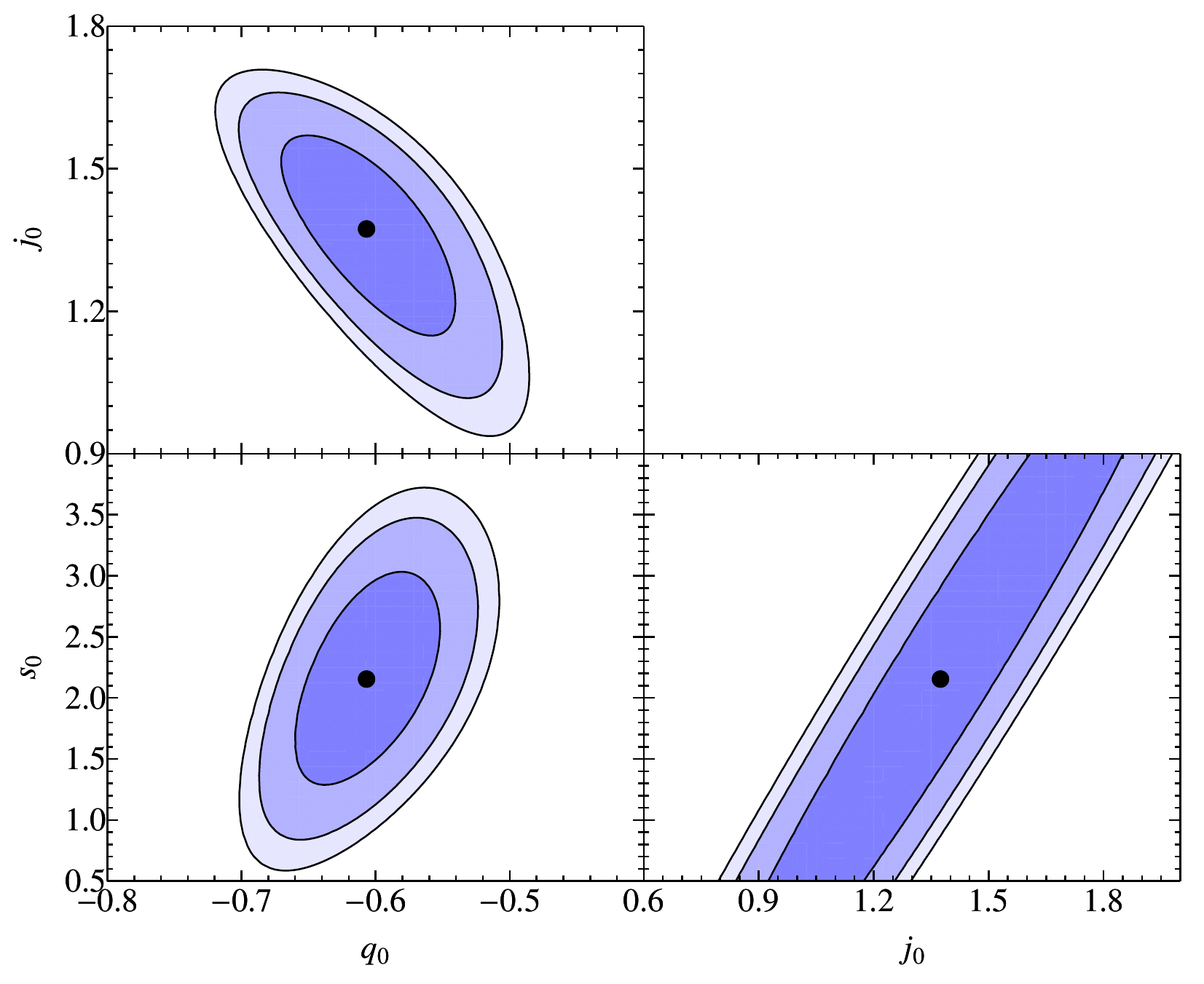}
\caption{$j_0$--$q_0$ (top panel), $s_0$--$q_0$ (bottom left panel), and $s_0$--$j_0$ (bottom right panel) contours of the cosmographic fit. The black dots and the blue shaded areas have the same meaning as in Fig.~\ref{figura2}.}
\label{figura3}
\end{figure*}

\noindent Cosmography is part of the set of model-independent procedures, which infer kinematical parameters from a direct fit of the Taylor series of the luminosity distance\footnote{See \cite{applications} for applications of cosmography in background cosmology.}. The advantage of this procedure is that it uses a very few assumptions.  In particular, considering  homogeneity and isotropy, one can expand the luminosity distance $d_L$ and measure the coefficients entering the expansion \cite{wei,star1,visser1,visser2,visserb}.
In particular, up to the fourth order in $z$ we have
\begin{equation}
\label{zeriez}
d_L(z) = d_H z \left[ 1 + c_1 z + c_2 z^2 + c_3 z^3 + \mathcal{O}(z^4) \right]\,,
\end{equation}
where $d_H \equiv c/H_0$ and
\begin{align}
c_1=&\frac{1}{2}\left(1-q_0\right)\quad,\quad c_2=-\frac{1}{6}\left(2c_1-3q_0^2+j_0+\Omega_k \right)\,,\nonumber\\
c_3=&-\frac{1}{24}\left[(30c_2+10c_1-\Omega_k)(1+q_0)+4(\Omega_k-c_1)-s_0\right].\nonumber
\end{align}
Once the expansion for $d_L$ is evaluated, similar expansions can be evaluated for the scale factor $a(t)$, the pressure, the density and so on.
The numerical coefficients of the Taylor expansion are related to the deceleration parameter $q_0$, the jerk parameter
$j_0$ (the variation of the acceleration), and the snap parameter $s_0$ (the rate of variation of the acceleration).
These parameters fully characterize the universe kinematics and are related to the free
parameters of a given model through their definitions:
\begin{align}\label{ciao}
q(t)&=-\frac{\dot{H}}{H^2} -1\,, \nonumber\\
j(t)&=\frac{\ddot{H}}{H^3}-3q-2\,,\nonumber\\
s(t)&=\frac{\dddot{H}}{H^4}+4j+3q(q+4)+6\,.\nonumber
\end{align}
Cosmography provides accurate results at small redshifts, but is affected  by severe convergence problems, truncation of Taylor series and error propagation, which cannot provide  accurate fitting outcomes. Different formulations of Eq.~\eqref{zeriez} have been presented to heal the above caveats \cite{visserb,trest,giovanni}. Once rewritten, the luminosity distance can be used to fit the SNe Ia data from Union $2.1$.
We fix the curvature parameter to the value $\Omega_k=0.001\pm0.002$ obtained by the \textit{Planck} mission \cite{Planck2018}.
The best-fit value of the parameters $q_0$, $j_0$, and $s_0$ and the value of the $\chi^2$ over the DOF  are summarized in the top part of Tab.~\ref{table:no3}. The contour plots of the best-fit parameters are shown inD Fig.~\ref{figura3}.
Inverting the results of the first three columns of Tab.~\ref{table:no3}, we obtain constraints on the free parameters of the HL model by using their mathematical correspondence with the cosmographic set which can be expressed as
\begin{eqnarray}
\label{q0w0}
q(z)&=&\{\Omega_{k}^{2}(1+z)^4+2\Omega_{\Lambda}\left[f(z)-3\Omega_\Lambda\right]\}/[4\epsilon(z)]\,,\\
\label{j0w0}
j(z)&=&\{3\Omega_{k}^{2}(1+z)^4+4\Omega_\Lambda f(z)\}/[4\epsilon(z)]\,,\\
\nonumber
s(z)&=&\{48\Omega_{k}\left(\Omega_\Lambda-\Omega_\omega\right)(1+z)^5g(z)+15\Omega_{k}^{4}(1+z)^8+\\
\label{s0w0}
&{}&-8\Omega_\Lambda^{2}[4f(z)+3\Omega_m(1+z)^3]\}/[4\epsilon(z)]\,,
\end{eqnarray}
where
\begin{align}
\nonumber
f(z)\equiv&\Omega_\Lambda+\Omega_m(1+z)^3\,,\\
\nonumber
g(z)\equiv&\Omega_m\Omega_\Lambda+\Omega_{k}^2(1+z)\,,\\
\nonumber
\epsilon(z)\equiv&\Omega_{k}^2(1+z)^4+\Omega_{k}\left(\Omega_\Lambda-\Omega_\omega\right)(1+z)^2+\Omega_\Lambda f(z)\,.
\end{align}
For today's value $z=0$, we obtain
\begin{eqnarray}
\label{1}
q_0&=&\left[\Omega_{k}^{2}-2\Omega_\Lambda\left(2\Omega_\Lambda-\Omega_m\right)\right]/(4\epsilon_0)\,,\\
j_0&=&\left(3\Omega_{k}^{2}+4\Omega_\Lambda f_0\right)/(4\epsilon_0)\,,\\
s_0&=&\left[48\Omega_{k}\left(\Omega_\Lambda-\Omega_\omega\right)g_0+15\Omega_{k}^{4}+\right.\nonumber\\
&&\left.-8\Omega_{\Lambda}^{2}(4f_0+3\Omega_m)\right]/(4\epsilon_0)\,,
\end{eqnarray}
with $f_0\equiv\Omega_\Lambda+\Omega_m$, $g_0\equiv\Omega_m\Omega_\Lambda+\Omega_{k}^2$, and $\epsilon_0\equiv\Omega_{k}^2+\Omega_{k}\left(\Omega_\Lambda-\Omega_\omega\right)+\Omega_\Lambda f_0$.

\noindent The bounds of Eq.~\eqref{jew} reduce the number of independent HL parameters to three.
It is, therefore, possible  to derive numerical bounds for  $\Omega_\omega$ in terms of the values of $q_0,j_0,$ and $s_0$.
Inverting Eqs.~\eqref{1}, we have\footnote{We do not consider $s_0$ since its value is affected by high systematic errors.}
\begin{eqnarray}
\label{inversione}
\Omega_\omega(q_0)&=&\Omega_\Lambda+\frac{2\Omega_\Lambda(\Omega_\Lambda+f_0)+(\Omega_{k}^{2}+4\Omega_\Lambda f_0)(q_0-1)}{4\Omega_{k}q_0}\,,\nonumber\\
\nonumber
\Omega_{\omega}(j_0)&=&\Omega_\Lambda-\frac{\Omega_k}{2}+\frac{(3\Omega_{k}^{2}+\Omega_\Lambda f_0)(j_0-1)}{4\Omega_k j_0}\,,
\end{eqnarray}
The numerical bounds on $\Omega_\omega(q_0)$ and $\Omega_{\omega}(j_0)$, obtained by using $\Omega_m=0.3153\pm0.0073$ and $\Omega_\Lambda=0.6847\pm0.0073$ from the \textit{Planck} mission \cite{Planck2018}, are summarized in the bottom part of Tab.~\ref{table:no3}.

\noindent In the $\Lambda$CDM limit, instead we easily get
\begin{eqnarray}
\label{inv1}
\Omega_m\left(q_0\right)&=&\frac{2}{3}\left(1+q_0\right)\,,\\
\label{inv2}
\Omega_m\left(q_0,j_0\right)&=&\frac{2}{3}\left(j_0+q_0\right)\,.
\end{eqnarray}
Their value are summarized in the last two column of the bottom part of Tab.~\ref{table:no3}.
\begin{table}[t]
\label{table:no3}
\begin{center}
\begin{tabular}{c|c|c|c}
\hline\hline
\multicolumn{4}{c}{Best-fit parameters}\\
\hline
$q_0$							& $j_0$											& $s_0$						& $\chi^2$/DOF	\\
\hline
$-0.607\pm0.066$	& $1.39\pm0.63$							& $2.2\pm3.2$			& $562.20/577$	\\
\hline\hline
\multicolumn{4}{c}{Derived quantities}\\
\hline
$\Omega_\omega(q_0)$	& $\Omega_\omega(j_0)$	& $\Omega_m(q_0)$ & $\Omega_m(q_0,j_0)$ \\
\hline
$129\pm361$					& $67\pm163$ 				& $0.262\pm0.044$ & $0.51\pm0.42$       \\
\hline
\end{tabular}
\caption{Top part: cosmographic best-fit values of $q_0$, $j_0$, and $s_0$ and the statistic test $\chi^2$/DOF. Bottom part: the derived bounds on $\Omega_\omega$ obtained from $q_0$ and $j_0$ and the results for $\Omega_m$ in the $\Lambda$CDM limit as described in Eqs.~\eqref{inv1}--\eqref{inv2}.}
\end{center}
\end{table}

\begin{figure*}
\centering
\includegraphics[width=16cm,clip]{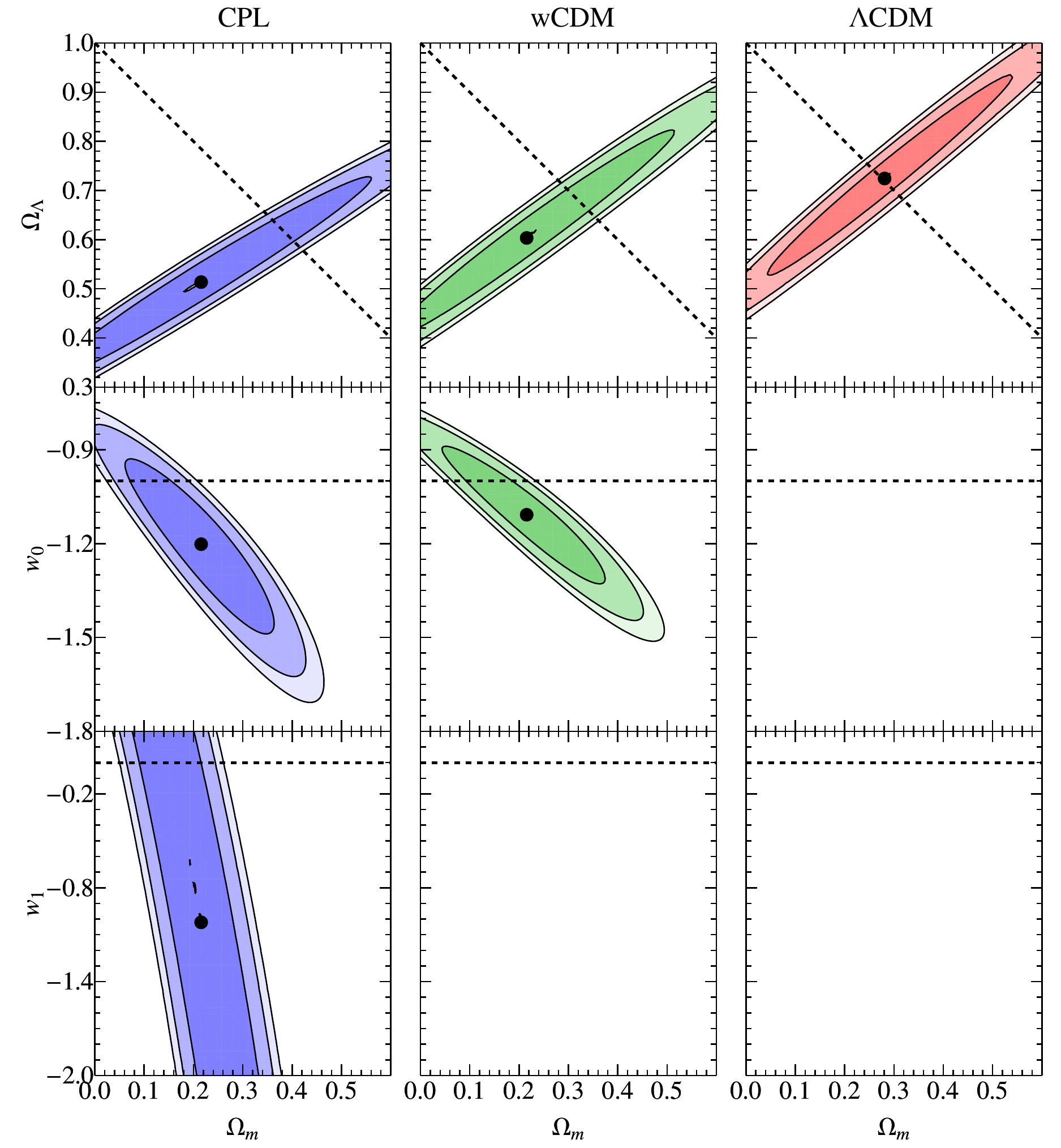}
\caption{Left column: $\Omega_\Lambda$--$\Omega_m$ (top panel), $w_0$--$\Omega_m$ (middle panel), and $w_1$--$\Omega_m$ (bottom panel) contours of the CPL model. Middle column: $\Omega_\Lambda$--$\Omega_m$ (top panel), and $w_0$--$\Omega_m$ (middle panel) contours of the wCDM model. Right column: $\Omega_\Lambda$--$\Omega_m$ (top panel) contour of the $\Lambda$CDM model. The black dots mark the best-fit values and the shaded areas (blue for the CPL, green for the wCDM, and red for the $\Lambda$CDM) show the $1$, $2$, and $3$--$\sigma$ confidence regions (from the darker to the lighter). The dashed lines in the top panels mark the allowed values of $\Omega_\Lambda$ and $\Omega_m$ for a flat universe.}
\label{figura4}
\end{figure*}


\begin{table*}[htb!]
\label{table:no4}
\begin{center}
\begin{tabular}{c|c|c|c|c|c|c|c|c|c|c}
\hline\hline
Model 				& $\Omega_m$			& $\Omega_\Lambda$	& $w_0$						&	$w_1$					&	$\Omega_k$			& $\Omega_\omega$	&	
$\chi^2$/DOF	&	$k$							& $\Delta$AIC 			& $\Delta$BIC 		\\
\hline
$\Lambda$CDM	& $0.28\pm0.25$		& $0.73\pm0.20$ 		& 								&								&									& 								&	
$562.23/578$	&	$2$ 						&	$0$								& $0$							\\
\hline
wCDM					& $0.22\pm0.34$		& $0.61\pm0.25$			&	$-1.11\pm0.18$	&								& 								&									&
$562.22/577$	& $3$ 						& $1.99$						& $6.36$					\\
\hline
CPL						& $0.22\pm0.34$		& $0.52\pm0.21$			&	$-1.2\pm0.27$		&	$-1.0\pm2.7$	& 								&									&
$562.20/576$	& $4$ 						& $3.97$						& $12.69$					\\
\hline
Ho\v rava			& $0.285\pm0.094$	& $0.729\pm0.042$		&									&								& $-0.02\pm0.12$	&	$0.08\pm3.68$		&
$562.23/576$	& $4$ 						& $4.00$						& $12.73$					\\
\hline
\end{tabular}
\caption{The best-fit parameters of the $\Lambda$CDM, wCDM, CPL, and HL models, the corresponding values of the $\chi^2$ over the DOF, the number of model parameters $k$, and the results of the statistical tests for BIC and AIC. The number of data points in all tests is $N=580$.}
\end{center}
\end{table*}

\section{Comparing Ho\v rava-Lifshitz cosmology with other models}

To compare different cosmological models, we will use statistical tests \cite{AIC,trs,do,qua,quu,tro} of  the Akaike Information
Criterion (AIC) and the Bayesian Information Criterion (BIC). These criteria are used as model-independent statistical methods for comparing different models \cite{trn}. They represent standard diagnostic tools of regression models
\cite{do,qua,quu,tro}. The formulation of the AIC criterion makes use of two distribution functions: $f(x)$ is postulated to be the \emph{exact} function describing a particular physical phenomenon; $g(x|\theta)$ approximates $f(x)$ through a set of parameters $\theta$.
By construction, there is only \emph{one set}, say $\theta^\star$, which minimizes the difference $|g(x|\theta)-f(x)|$.
Evaluating the AIC value for a single model is clearly unfeasible, since $f(x)$ is unknown \emph{a priori}.
For a Gaussian error distribution, the generic AIC, calculated over the whole set of models, is given by
\begin{equation}
{\rm AIC} = \chi^2 + 2 k\,,
\end{equation}
where $k$ is the number of model parameters.
Hence, any speculations may be computed on the basis of
\begin{equation}
\label{dell}
\Delta{\rm AIC} \equiv \Delta\chi^2 + \Delta k\,,
\end{equation}
where each $\chi^2$ is computed for the set $\theta^\star$ of each model.

\noindent The BIC criterion has been derived in a Bayesian context \cite{qud,qut} and is defined as
\begin{equation}
{\rm BIC}=\chi^2+k\ln N\,,
\end{equation}
where $N$ is the number of data points used throughout the fit procedure.
The difference in BIC is given by
\begin{equation}
\Delta{\rm BIC}=\Delta\chi^2+\Delta k\ln N.
\end{equation}
We employ the AIC and BIC techniques to compare the HL cosmology with the $\Lambda$CDM model, the wCDM model, which is the natural extension of the $\Lambda$CDM paradigm and provides a negative barotropic factor inside the interval $w_0\geq-1$, and the varying quintessence model proposed by the Chevallier, Polarsky and Linder (CPL parametrization) \cite{cipielle}, based on the Taylor series of the
EoS: $w(a)=w_0+w_1(1-a)$.

\noindent The three last models are characterized, respectively, by
\begin{subequations}
\label{Ediz}
\begin{align}
E_{\Lambda{\rm CDM}}&=\sqrt{\alpha(z)+\Omega_\Lambda}\,,\\
E_{\rm wCDM}&=\sqrt{\alpha(z)+\Omega_\Lambda\left(1+z\right)^{3(1+w_0)}}\,,\\
E_{\rm CPL}&=\sqrt{\alpha(z)+\Omega_\Lambda\tau(z)}\,,
\end{align}
\end{subequations}
where $\alpha(z)=\Omega_m\left(1+z\right)^3+(1-\Omega_m-\Omega_\Lambda)(1+z)^2$ and $\tau(z)=(1+z)^{3(1+w_0+w_1)}\exp\left[-3 w_1 z/(1+z)\right]$.
The contour plots of the fits performed with the $\Lambda$CDM, wCDM and CPL models are shown in Fig.~\ref{figura4}.
The best-fit parameters of the above models and of the HL model together with the results of the AIC and BIC tests are summarized in Tab.~\ref{table:no4}.

\noindent These results show definitively that the HL cosmological is:

\begin{itemize}
  \item statistically disfavored if compared to the $\Lambda$CDM approach;
  \item disfavored even if compared with the CPL parametrization. In this case, the two models have the same number of parameters;
  \item kinematics does not favor the Ho$\check{\textrm{r}}$ava model and are compatible with statistics.
\end{itemize}

\noindent Thus, the HL cosmology is both kinematically and statistically disfavored with respect to the standard cosmological scenarios represented by the $\Lambda$CDM, wCDM and CPL models. Moreover, the HL model does not properly pass the cosmographic limits. This fact is supported by the AIC and the BIC,  which show that, independently of the number of free parameters, the HL cosmology is not consistent with observational data.

\section{Final outlooks}

In this paper, we investigated the limits imposed by observations on the minimal paradigm of HL cosmology in a homogeneous and isotropic background. To do so, we first consider kinematics as a model-independent approach to check whether the HL cosmology is in accordance with the SN Ia data surveys. By using the approach of cosmography, we found viable intervals for the values of the free parameters entering the model and then we compared these results with fits performed by using the Union 2.1 compilation. The $1\sigma$ agreement with
cosmography is quite acceptable and leaves open the possibility that the model works  well in describing large scale dynamics.
Although appealing, the numeric outcomes contradict the statistical requirements that come from the two selection criteria we use here,   i.e.,  the AIC and BIC statistics. We found that the corresponding $\Delta$AIC and $\Delta$BIC functions are wide under the numerical constraints previously obtained from cosmography and cosmic fits. This suggests that the HL model is less accurate than the standard $\Lambda$CDM paradigm. If one enlarges the number of free parameters within the cosmic scenario, including the wCDM and CPL frameworks, the HL cosmology continues being disfavored. This result points out the statistical inconsistencies of the model, ruling out \emph{de facto} the HL paradigm at the level of background cosmology. Future analysis could focus on verifying the validity of the HL cosmology at both early and late times. This may be useful to find viable HL models, by adding new extra terms into the Lagrangian.

\section*{Acknowledgments}

This work was partially supported  by UNAM-DGAPA-PAPIIT, Grant No. 111617, and by the Ministry of Education and Science of RK, Grant No.
BR05236322 and AP05133630.

\end{document}